\documentclass[preprint]{elsarticle}
\usepackage{amsmath,amssymb}
\usepackage{lscape}

\begin{document}
\title{An experiment of X-ray photon-photon elastic scattering with a Laue-case beam collider}
\author[phys]{T.~Yamaji}
\ead{yamaji@icepp.s.u-tokyo.ac.jp}
\author[icepp]{T.~Inada}
\author[icepp]{T.~Yamazaki}
\author[icepp]{T.~Namba}
\author[phys]{S.~Asai}
\author[kek]{T.~Kobayashi}
\author[sp8]{K.~Tamasaku}
\author[hyogo]{Y.~Tanaka}
\author[jasri]{Y.~Inubushi}
\author[sp8]{K.~Sawada}
\author[sp8]{M.~Yabashi}
\author[sp8]{T.~Ishikawa}
\address[phys]{Department of Physics, Graduate School of Science, The University of Tokyo, 7-3-1 Hongo, Bunkyo, Tokyo 113-0033, Japan}
\address[icepp]{International Center for Elementary Particle Physics, The University of Tokyo, 7-3-1 Hongo, Bunkyo, Tokyo 113-0033, Japan}
\address[kek]{High energy Accelerator Research Organization, KEK, 1-1 Oho, Tsukuba, Ibaraki 305-0801, Japan}
\address[sp8]{RIKEN SPring-8 Center, 1-1-1 Kouto, Sayo-cho, Sayo-gun, Hyogo 679-5148, Japan}
\address[hyogo]{Graduate School of Material Science, University of Hyogo, 3-2-1 Kouto, Kamigori-cho, Ako-gun, Hyogo 678-1297, Japan}
\address[jasri]{Japan Synchrotron Radiation Research Institute (JASRI), 1-1-1 Kouto, Sayo-cho, Sayo-gun, Hyogo 679-5198, Japan}

\begin{abstract}
We report a search for photon-photon elastic scattering in vacuum in the X-ray region at an energy in the center of mass system of $\omega_{\rm cms} =6.5$ keV for which the QED cross section is $\sigma_\mathrm{QED} =2.5 \times 10^{-47} \ \mathrm{m^2}$. An X-ray beam provided by the SACLA X-ray Free Electron Laser is split and the two beamlets are made to collide at right angle, with a total integrated luminosity of  $(1.24 \pm 0.08) \times 10^{28} {\rm \ m^{-2}}$.
No signal X rays from the elastic scattering that satisfy the correlation between energy and scattering angle were detected. We obtain a 95\% C.L. upper limit for the scattering cross section of $1.9\times 10^{-27}\ \mathrm{m^2}$ at $\omega_\mathrm{cms}=6.5\ \mathrm{keV}$. The upper limit is the lowest upper limit obtained so far by keV experiments.
\end{abstract}

\maketitle

\section{Introduction}
Within the framework of classical electrodynamics, light cannot interact with light. However, quantum electrodynamics (QED) predicts that vacuum polarization, a nonlinear effect of quantum fluctuations, intermediates photon-photon elastic scattering. A theoretical cross section was first calculated in 1933 using the Dirac theory \cite{1}, and was later done using QED \cite{2}. The leading contribution to the photon-photon scattering cross section is described by a fourth-order Feynman diagram with an electron-positron loop (a box diagram),  whose scattering amplitude in the low energy region is strongly suppressed by the electron loop.

\vspace{\baselineskip}

When the photon energy in the center of mass system  ($\omega_{\rm cms}$) is less than $700\ {\rm keV}$, the first-order QED cross section for photons  with the same linear polarization state can be described as 
\begin{eqnarray}
%\left(\frac{d\sigma_{\gamma\gamma\rightarrow\gamma\gamma}}{d\Omega}\right)_{\rm QED}&=& \frac{\alpha^4 \omega_{\rm cms}^6}{(180\pi)^2m_{\rm e}^8}(260\mathrm{cos}^4\theta+328\mathrm{cos}^2\theta+580),\\
\left(\frac{d\sigma_{\gamma\gamma\rightarrow\gamma\gamma}}{d\Omega}\right)_{\rm QED}&=& \frac{\alpha^2 r_{\rm e}^2}{(180\pi)^2}\left(\frac{\omega_{\rm cms}}{m_{\rm e} c^2}\right)^6(260\mathrm{cos}^4\theta+328\mathrm{cos}^2\theta+580),\\
\sigma_\mathrm{QED}&=&3.5\times 10^{-70}(\omega_{\rm cms} \ [\mathrm{eV}])^6\ \mathrm{m^2},
\end{eqnarray}
where $\alpha$ is the fine structure constant, $r_{\rm e}$ is the classical electron radius, and $\theta$ is the scattering angle between an incident photon and a scattered photon in the center of mass system ($0 \le {\rm cos}\theta \le 1$, $\theta=0$ corresponds to forward scattering) \cite{2}. The cross section is proportional to the sixth power of $\omega_{\rm cms}$.

\vspace{\baselineskip}

Although the inclusive contribution of a box diagram with virtual photons in the MeV region is observed by Delbr\"{u}ck scattering \cite{delbruck} and the high-precision measurement of electron and muon $g-2$ \cite{high-pre}, the process with real photons has not ever been directly observed. The observation of photon-photon scattering of real photons would provide solid evidence for vacuum polarization caused by virtual electrons. Furthermore, the search for elastic scattering of real photons has a particular interest as a search for new physics. The photon-photon scattering cross section can be enhanced since new particles predicted by physics beyond the standard model may mediate photon-photon scattering. For example, Axion Like Particles (ALPs), pseudoscalar bosons which have a two-photon coupling constant uncorrelated to their mass, can mediate photon-photon scattering by {\it s} or {\it t}-channel virtual exchange of ALPs \cite{7,8,9}. 

%For example, the axion, a pseudoscalar boson, can mediate photon-photon scattering by {\it s} or {\it t}-channel virtual axion exchange\cite{7,8,9}. 

\vspace{\baselineskip}

Previous searches have been performed using high intensity optical or infrared lasers \cite{5,6}. However, their sensitivity to photon-photon scattering is suppressed by the tiny QED cross section in the low energy region, $\mathcal{O}(10^{-70})\ \mathrm{m}^2$ at 1 eV, and by white background photons  generated in their optical systems. By using X rays, the photon-photon cross section can be enhanced by around 23 orders of magnitude compared to the optical region and background photons can be reduced by high precision energy measurement. We performed photon-photon scattering experiments with $\omega_{\rm cms}$ of 6.5 keV and $\sigma_\mathrm{QED}$ of $2.5 \times 10^{-47} \ \mathrm{m^2}$. 
%Furthermore, signal X rays can be distinguished from background photons by using the energy information. 

\vspace{\baselineskip}

In 2013 we performed the first photon-photon scattering experiment in the X-ray region at the Spring-8 Angstrom Compact free-electron LAser (SACLA) \cite{SACLA}. No signal was observed, and we obtained an upper limit on the cross section of photon-photon scattering of $\sigma_{\gamma\gamma\rightarrow\gamma\gamma} < 1.7 \times 10^{-24}{\rm \ m^2}$ (95\% C.L.) at $\omega_{\rm cms}=6.5{\rm \  keV}$ \cite{10}. 

\vspace{\baselineskip}

In this paper, we report on the result of a recently upgraded experiment at SACLA with an optimized setup. In this experiment, we replaced the X-ray beam collider with one with thinner blades. This yields a higher diffraction efficiency and smaller sizes of  the colliding beams. The beam quality from SACLA has also been improved, with respect to the repetition rate, bandwidth and pulse intensity.

\section{Experimental setup and measurement}
SACLA generated horizontally-polarized X-ray pulses with $\sim10^{11}$ photons per pulse, a repetition rate of 30 Hz, a bandwidth of about 50 eV (FWHM), a beam width of 200 $\mathrm{\mu m}$ (FWHM), and a pulse duration less than 10 fs (FWHM) \cite{11}. The vertical and horizontal Rayleigh lengths of X-ray pulses are of the order of 100 m.  In this experiment, the energy of the initial X ray was set to 10.985 keV. For extracting X rays with energy within the acceptable bandwidth of our beam collider ($100\mathrm{\ meV}$), X-ray beams were monochromatized to $60\mathrm{\ meV}$ by two silicon (440) channel-cut monochromators. The pulse intensity was reduced by a monochromatic factor of $60\mathrm{\ meV}/50\mathrm{\ eV}\simeq 10^{-3}$, and the photon number of the monochromatic beams was around $10^{8}$ photons per pulse. The pulse duration of the monochromatic beams was expanded to $\sim 30$ fs due to the time-energy uncertainty relation. The monochromatic beams were focused to $\sim 1\mathrm{\ \mu m}$ (FWHM) in the horizontal direction by an elliptical mirror \cite{13}, and the horizontal Rayleigh length is reduced to 20 mm. A detailed explanation of the beamline optics is given in our previous paper \cite{10}.

\vspace{\baselineskip}

\begin{figure}
  \centering
  \includegraphics[width=0.9\textwidth]{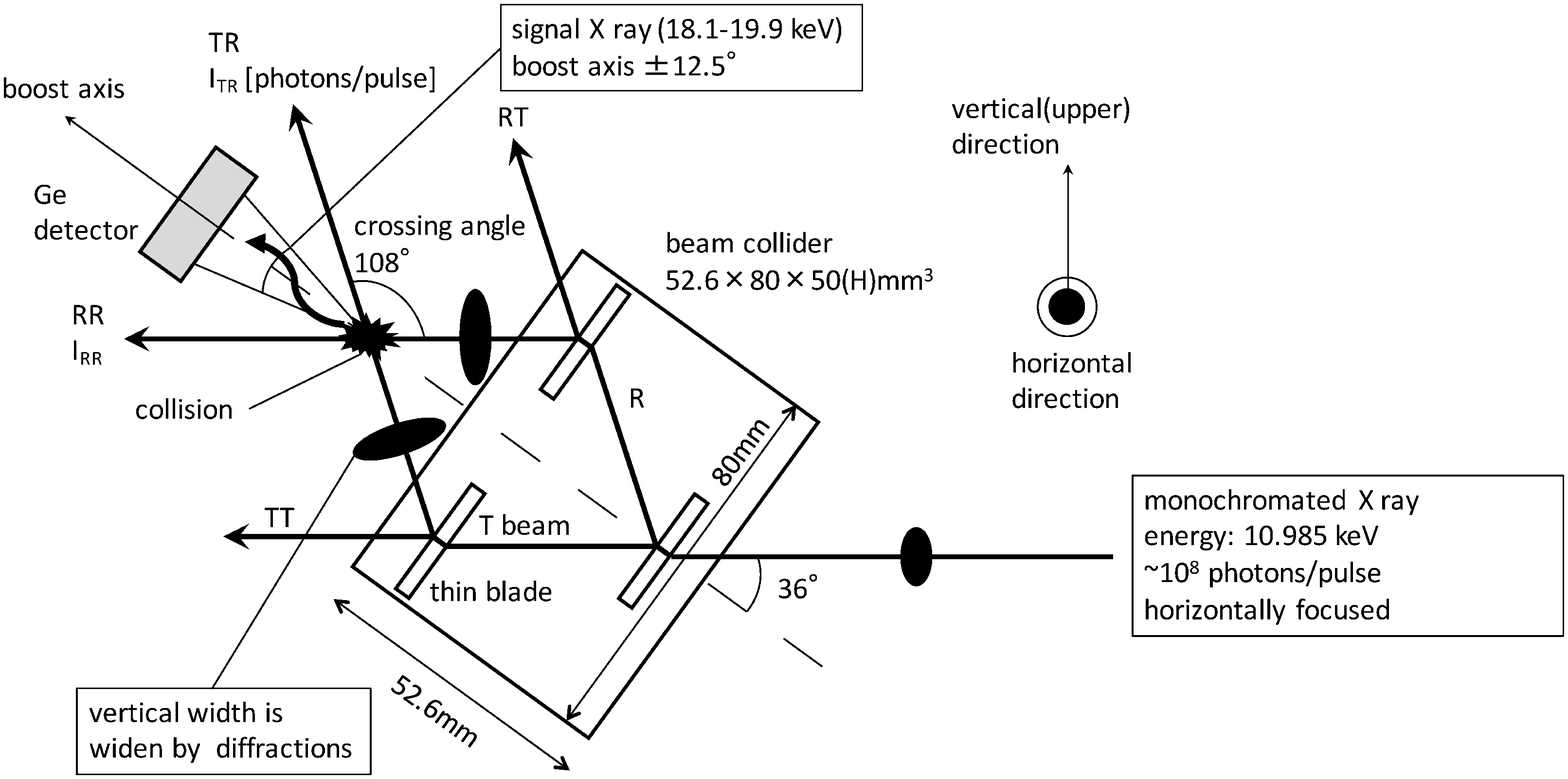}
  \caption{The schematic of the collision system. Vertical and horizontal directions, which is normal to the optical axis, are also shown.}
  \label{bm_schematics}
\end{figure}

A schematic of the collision system using an X-ray beam collider is shown in Fig. \ref{bm_schematics}. The X-ray beam collider, composed of three thin blades manufactured on a silicon single crystal, diffracts X-ray beams in the transmission (Laue) geometry with the (440) plane at the Bragg angle of $36^\circ$ \cite{14}. The X-ray beam collider divides the incident beam into four beams (RR, RT, TR, TT beams shown in Fig. \ref{bm_schematics}), two of which (RR/TR) collide obliquely with a crossing angle of $108^\circ$ and $\omega_{\rm cms}$ of $6.5\mathrm{\ keV}$. The collision is assured spatially and temporally since the lengths of light paths of RR and TR beams are geographically the same  \cite{14}. The diffraction efficiencies are higher for the collider with thinner blades because diffracted X rays suffer smaller X-ray absorption within blades. The thickness of the blades is 0.2 mm, and the measured diffraction efficiencies of the colliding beams are 1.55\% (RR) and 2.48\% (TR). The beam collider was installed in a vacuum chamber evacuated at a pressure of less than $10^{-2}$ Pa in order to reduce stray photons scattered by the residual gas.  The alignment of the chamber in the optical axis was performed by using a manual stage with the precision of 1 mm, much shorter than the Rayleigh lengths. The vertical and horizontal positions, which is normal to the optical axis, were aligned by scanning motorized stages and monitoring the intensity of TR/RR beams passing through the chamber with the precision of less than 0.1 mm.  

\vspace{\baselineskip}

Since the center of mass system of the two colliding photons is boosted with a Lorentz factor of $\gamma$=1.7, the spatial distribution of signal X rays is concentrated along the boost axis, and the scattered X rays on the boost axis have higher energy than the incident X ray. A germanium semiconductor detector (CANBERRA BE2825) was used to measure the energy of the scattered X rays \cite{10}. The energy and time resolution of the detector were measured to be $200\mathrm{\ eV}$ ($1\sigma$, for 26.3-keV X rays) and $80\mathrm{\ ns}$ ($1\sigma$), respectively. The detector was located on the boost axis of the experimental system as shown in Fig. \ref{bm_schematics}, and detected scattered X rays within a cone with an apex angle of $25^\circ$ around the boost axis. Within this cone, the signal X rays have an energy between 18.1 and 19.9 keV, and the signal coverage is 17\%. 

\vspace{\baselineskip}

The detector was operated with external triggers synchronized to SACLA. To reject environmental background, a time window was set to $\pm 0.4 \mathrm{\ \mu s}$, which corresponds to $\pm 5\sigma$ of the detector time resolution. The energy was required to be between $17.6$ and $20.4 \mathrm{\ keV}$, widened from the original signal energy range by the detector energy resolution ($\pm2\sigma$). The detection efficiency of the detector is $\epsilon =(13.2\pm 0.3 $)\%, as estimated by GEANT4 simulation, and cross-checked by using X-ray sources of $^{55}$Fe, $^{57}$Co, $^{68}$Ge, and $^{241}$Am \cite{10}. The uncertainty on the absolute detection efficiency was taken as a systematic uncertainty. 

\vspace{\baselineskip}

The single-pulse integrated luminosity for obliquely colliding Gaussian beams without angular divergences ($L_{\rm pls}$) is given by the following formula \cite{15}, 
\begin{eqnarray}
L_{\rm pls}=\frac{I_{\rm RR}I_{\rm TR}}{4\pi \sigma_{\rm h} \sigma_{\rm v}\sqrt{1+\frac{\sigma_{\rm t}^2}{\sigma_{\rm v}^2}\mathrm{tan^2}\left(\frac{\theta_{\rm c}}{2}\right)}} \simeq \frac{I_{\rm RR}I_{\rm TR}}{4\pi \sigma_{\rm h}\sigma_{\rm v}},
\end{eqnarray}

where $I_{\rm RR/TR}$ are the photon number of colliding beams, $\sigma_{\rm h/v/t}$ are the beam widths (1$\sigma$) in the horizontal, vertical, and  time directions, respectively, and $\theta_{\rm c}=108^\circ$ is the crossing angle. The crossing duration of two beams is $\sim \sigma_{\rm v}/c$ in the oblique collision. Although RR/TR beams have angular divergencies, the uncertainty of the kinematic factor and the luminosity-reduction effect (Hourglass effect) arising from the angular divergencies are negligible because the angular divergences are small ($1\mathrm{\ \mu rad}$/$40\mathrm{\ \mu rad}$ for the vertical/horizontal direction) and the vertical and horizontal Rayleigh lengths (100 m and 20 mm) are much longer than the length of luminous region on the optical axis ($\sim \sigma_{\rm v}$). The reduction factor $\sqrt{1+\frac{\sigma_t^2}{\sigma_v^2}\mathrm{tan^2}\left(\frac{\theta_c}{2}\right)}$, which comes from the oblique collision, is approximated to be 1 since $\sigma_{\rm t}$ ($\sim 30{\ \rm fs}=9{\rm \ \mu m}$) is much smaller than $\sigma_{\rm v}$  ($\sim 100\mathrm{\ \mu m}$). When the vertical or horizontal widths of the RR and TR beams ($\sigma_{\rm v}^{\rm RR/TR}$, $\sigma_{\rm h}^{\rm RR/TR}$) are different, effective values of $\sigma_{\rm h,v}$ are estimated as $\sigma_{\rm h,v}=\sqrt{\frac{(\sigma_{\rm h,v}^{\rm RR})^2+(\sigma_{\rm h,v}^{\rm TR})^2}{2}}$.

\vspace{\baselineskip}

The horizontal and vertical widths of the colliding beams were measured by knife edge scans using a gold wire ($\phi=200\ \mathrm{\mu m}$) and a steel rod, respectively. We assume that the colliding beams have a Gaussian profile, and the widths were measured by scanning the edge and measuring the transmitted intensity using silicon PIN photodiodes (PD, HAMAMATSU S3590-09). The transmitted intensity is well-fitted by an error function within the fluctuation of the beam intensity, and the horizontal and vertical widths ($1\sigma$) were measured to be  
$\sigma_{\rm h}^{\rm RR}=(0.962\pm 0.037) \mathrm{\ \mu m}$ and $\sigma_{\rm h}^{\rm TR}=(0.992\pm 0.044) \mathrm{\ \mu m}$ and $\sigma_{\rm v}^{\rm RR}=(143\pm 12) \mathrm{\ \mu m}$ and $\sigma_{\rm v}^{\rm TR}=(124\pm 7)\mathrm{\ \mu m}$, respectively. These uncertainties on beam sizes, which are considered as a source of  a systematic uncertainty, originate from fluctuations of SACLA intensity during the knife edge scans. From these measured widths, the effective beam widths $\sigma_{\rm h,v}$ are calculated to be $\sigma_{\rm h}=(0.977\pm0.028)\ \mathrm{\mu m}$ and $\sigma_{\rm v}=(134\pm 7)\ \mathrm{\mu m}$, respectively. The vertical widths of the colliding beams were widened by Laue-case X-ray diffraction, and are proportional to the thickness of the blades \cite{19}. The vertical widths expected from the theoretical calculation of the shape of diffracted beams are $140$ $\mathrm{\mu m}$ (RR) and $130$ $ \mathrm{\mu m}$ (TR), consistent with the measurements. 

\vspace{\baselineskip}

The experiment was performed during a 34-hour DAQ time at SACLA in Nov. 2015, and a total of $3.7 \times 10^6$ X-ray pulses were used for the analysis. $I_{\rm RR}$ and $I_{\rm TR}$  were continuously monitored by PDs, and their mean product, calculated from the pulse-by-pulse product of $I_{\rm TR}I_{\rm RR}$, was $(5.53\pm 0.15) \times 10^{12} {\rm \ pulse^{-1}}$. The uncertainty on the detection efficiency of PDs calculated by a GEANT4 simulation was taken as a systematic uncertainty.

\vspace{\baselineskip}

There are two main background sources. The dominant background source is an accidental coincidence of environmental X rays. This background was measured in an off-timing window, and the accidental contribution was estimated to be 0.4 events in the energy range 17.6 - 20.4 keV.

\vspace{\baselineskip}

\begin{figure}
  \centering
  \includegraphics[width=0.9\textwidth, angle=0]{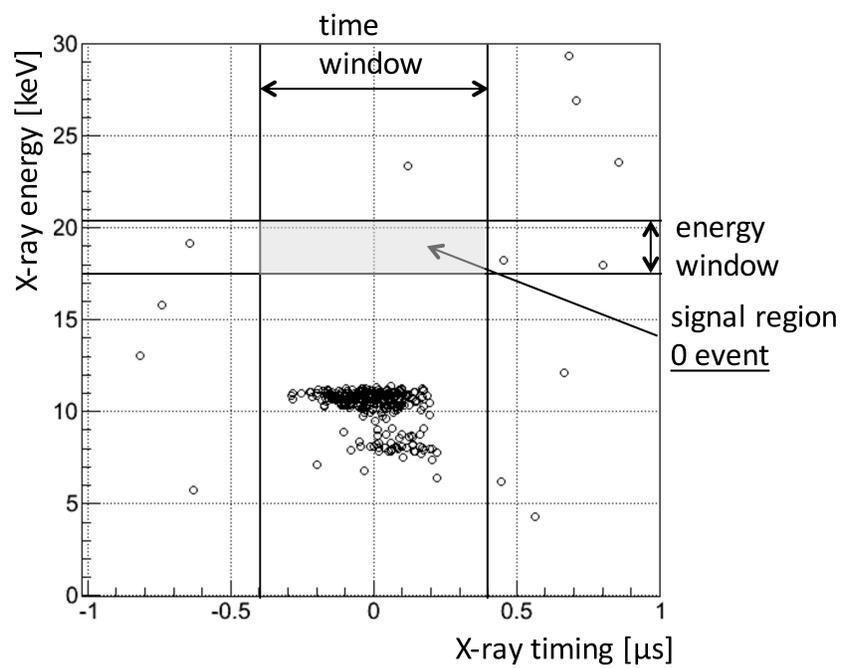}
  \caption{The distribution in time and energy of X rays measured with the germanium detector. The horizontal axis is the time with respect to the SACLA X-ray timing. Each circle represents an X-ray event.}
  \label{2d_spec}
\end{figure}

The measured distribution of X rays in time and energy is shown in Fig.\ref{2d_spec}. Events out of the time window are due to accidental coincidence of environmental X rays. Events within the time window and with energy less than 11 keV are stray photons from the X-ray beams, generated by Compton or Rayleigh scattering ($\sim 11$ keV) or X-ray fluorescence from chamber materials such as stainless steel ($\le 11$ keV). When two stray X rays coincide in the germanium detector, it may enter the signal region. This pileup is the second source of background.  The measured event rate of stray X rays is $(7.3\pm 0.4) \times 10^{-5}\ {\rm pulse^{-1}}$, and the expected pileup rate of two photons is estimated to be 0.01 events for $3.7\times 10^6$ pulses.

\vspace{\baselineskip}

No events were observed within the signal region, in which 0.4 background events were expected. By summing $L_{\rm pls}$ for all $3.7\times 10^6$ pulses, the total integrated luminosity $L_{\rm int}$ is calculated to be $(1.24 \pm 0.08) \times 10^{28} {\rm \ m^{-2}}$. The uncertainty on the total integrated luminosity is estimated by the root of the quadrature sum of relative uncertainties on $I_{\rm TR}I_{\rm RR}$ and $\sigma_{\rm h,v}$, assuming that these systematic uncertainties obey Gaussian distributions, and that they are independent. 

\vspace{\baselineskip}

The measured cross section of photon-photon scattering $\sigma_{\gamma\gamma\rightarrow\gamma\gamma}$ is estimated by
\begin{eqnarray}
\sigma_{\gamma\gamma\rightarrow\gamma\gamma}=\frac{\mu}{\epsilon L_{\rm int}},
\end{eqnarray}
where $\mu$ is the number of signal events. From this formula, the probability distribution function (PDF) of $\sigma_{\gamma\gamma\rightarrow\gamma\gamma}$ is calculated by combining the PDFs of $\epsilon$, $\mu$ and $L_{\rm int}$. Since no signal was observed in this experiment, the PDF of $\mu$ is an exponential distribution with a mean value of 1. We postulate that $\epsilon$ and $L_{\rm int}$ are normally distributed and independent of each other. Systematic uncertainties related to the upper-limit calculation are summarized in Table \ref{tab:sys}.

\vspace{\baselineskip}

From the cumulative distribution function of $\sigma_{\gamma\gamma\rightarrow\gamma\gamma}$, the 95\% C.L. upper limit is calculated to be 
\begin{eqnarray}
\sigma_{\gamma\gamma\rightarrow\gamma\gamma} < 1.9\times 10^{-27} {\rm \ m^2}
\end{eqnarray}
at $\omega_{\rm cms}=6.5{\rm \ keV}$. From the upper limit, the upper limit on the coupling constant of resonant ALPs to photons $g_{\rm lim}$ is calculated by 
\begin{eqnarray}
g_{\rm lim, resonant}=\left[16\frac{\sigma_{\gamma\gamma\rightarrow\gamma\gamma}}{(\hbar c)}\frac{\Delta \omega_{\rm cms}}{ \omega_{\rm cms}} \right]^{\frac{1}{2}}=2 \ {\rm GeV^{-1}}
\end{eqnarray}
at the ALPs mass of $6.5{\rm \ keV} \pm 25 {\rm \  meV}$, where $\Delta \omega_{\rm cms}=50\mathrm{\ meV}$ is the uncertainty of $\omega_{\rm cms}$ \cite{9}. The sensitivity for off-resonant ALPs with the mass of much lighter than $\omega_{\rm cms}$ is strongly reduced to $g_{\rm lim}=3.1\times 10^4 \ {\rm GeV^{-1}}$. Although this upper limit on the ALP coupling constant is much weaker than other limits obtained from laboratory experiments (see Figure 2 of \cite{alp_limit}), it provides the most stringent constraints so far with the direct scattering method in the X-ray region.
\vspace{\baselineskip}

%This result is 900 times more stringent than our previous upper limit. A breakdown of the enhancement factors is shown in Table \ref{tab:sum}. The main enhancement factors are the gains in the product of the intensities of the colliding beams $I_{\rm RR}I_{\rm TR}$ and the number of the measured pulses. $I_{\rm RR}I_{\rm TR}$ is enhanced by two orders of magnitude, which originates from the reduction of the bandwidth of SACLA (from 80 eV to 50 eV), the increase in the intensity of SACLA ($\sim1.8$ times higher), and the better diffraction efficiencies. In our previous experiment, we used an X-ray beam collider with blades  0.6mm thick. Meanwhile, the X-ray beam collider used in this experiment has thinner blades with the thickness of 0.2mm, which results in 11 times higher diffraction efficiencies, because diffracted X rays suffer smaller X-ray absorption within blades. The number of measured X-ray pulses was enhanced by a factor of 5.7, due to the increase in the repetition rate of SACLA (from 20 Hz to 30 Hz) and DAQ time (from 9 hours to 34 hours). 

\section{Conclusion}
We have searched for photon-photon scattering at SACLA with a Laue-case X-ray collider. No signal was observed in the expected signal region, and the obtained  95\% C.L. upper limit on the photon-photon scattering cross section was $\sigma_{\gamma\gamma\rightarrow\gamma\gamma} < 1.9\times 10^{-27} {\rm \ m^2}$ at $\omega_{\rm cms}=6.5{\rm \  keV}$. The upper limit is the lowest upper limit obtained so far by keV experiments, and that is 20 orders of magnitude above the QED cross section. 

\vspace{\baselineskip}

The sensitivity of our experiment is limited by the monochromatic factor ($\sim 10^{-3}$) and diffraction efficiencies ($\sim 1\%$). These small factors reduce the sensitivity, proportional to their square, by 10 orders of magnitude. It is technically difficult to enhance the efficiency of the current collision system, because the efficiency of the Laue-case diffraction is close to the limitation due to the crystal-cutting process.

\vspace{\baselineskip}

There are two possibilities to further enhance the sensitivity. First, a new method, self-seeding, can improve the monochromatic factor. The bandwidth of SACLA is currently limited to 50 eV because Self-Amplified Spontaneous Emission (SASE) \cite{16}, the amplification mechanism of XFEL, originates from the stochastic density fluctuation of the electron beams. Self-seeding will be introduced to SACLA in the near future, and will improve the bandwidth to $\sim$1 eV \cite{17}. This improved monochromaticity will reduce the monochromatic loss and enhance the integrated luminosity by around three orders of magnitude. 

\vspace{\baselineskip}

Secondly, a new collision system using a Bragg-case X-ray beam collider can enhance the diffraction efficiencies. For higher efficiency than the Laue-case beam collider used in this experiment, an X-ray beam collider using Bragg-case diffraction will enable $\sim 50\%$ efficiencies of colliding beams \cite{18}. In addition, Bragg diffraction does not expand the vertical widths of beams. By introducing a beam collider using Bragg-case diffraction, the integrated luminosity can be enhanced by around three orders of magnitude. 

\vspace{\baselineskip}

With these improvements, the integrated luminosity of our experiment in the X-ray region can be significantly enhanced by six orders of magnitude. In order to measure signal X rays in such a high-luminosity experiment, a new detection system which excludes the pileup of stray X rays is needed because the rate of pileups is proportional to the square of the intensity of monochromatized beams. In the high-luminosity experiment, the rate of pileups is $\mathcal{O}(10^3)$ times higher than this experiment. Position-sensitive detectors can identify piled up X rays by using the 2D position information of detected X rays and reduce the rate of pileups by the number of detection areas. A double-sided silicon strip detector (DSSD) is one candidate, which has $\mathcal{O}(10^3)$ active regions \cite{DSSD}.

\section{Acknowledgements}
The XFEL experiment was performed at BL3 in SACLA with the approval of JASRI and the Program Review Committee (Proposal No.  2014B8028 and 2015B8007). Before the experiment, preliminary experiments were performed at SPring-8 BL19LXU beamline with the proposal number 20140024/20150010.
We wish to acknowledge the help given by the engineering team of SACLA in the beamtime. The work of T. Yamaji is supported in part by the Advanced Leading Graduate Course for Photon Science (ALPS) at the University of Tokyo. This research was funded by the Japan Society for the Promotion of Science (Grant number 15J00509 and 15K17629), and MEXT (Grant number 26104701).
We would like to thank Daniel Jeans and Stefan Knirck for useful discussions and suggestions.

\clearpage

\clearpage

\clearpage

\begin{table}
  \begin{center}
    \begin{tabular}{|l|l|c|} \hline
      factor & origin & systematic uncertainties (relative, 1$\sigma$)\\ \hline \hline
      $I_{\rm RR}I_{\rm TR}$ & detection efficiencies of PDs & $\pm 2.7$\%\\ 
      $\sigma_{\rm h}$ & fluctuation of beam intensity & $\pm 2.9$\% \\ 
      $\sigma_{\rm v}$ & same as above & $\pm 5.2$\% \\ \hline
      $L_\mathrm{int}$ & $I_{\rm RR}I_{\rm TR},\sigma_{\rm h},\sigma_{\rm v}$ & $\pm 6.5$\% \\ \hline \hline
     $\epsilon$ & Geant4 simulation & $\pm 2.3$\% \\  \hline 
    \end{tabular}
     \caption{Systematic uncertainties related to the upper-limit calculation.}
    \label{tab:sys}
  \end{center}
\end{table}

\clearpage

 %\begin{table}
 %\begin{center}
 %   \begin{tabular}{|l|c|c|c||c|} \hline
 %     factor &  symbol & previous exp.(2013) & this exp.(2015) & gain \\ \hline \hline
 %     SACLA bandwidth $[\mathrm{eV}]$ & & $\sim$80 & $\sim$50 & 3 \\ 
 %     SACLA intensity $[\mathrm{arb}]$ & & 1 & $\sim$1.8 & 3 \\ 
 %     diffraction efficiencies (TR/RR) [\%] & & 0.63/0.54 & 2.48/1.55 & 11 \\ \hline
 %     $I_{\rm RR}I_{\rm TR}\ [\mathrm{pulse^{-1}}]$ &$I_{\rm RR}I_{\rm TR}$& $5.5\times 10^{10}$ & $5.5\times 10^{12}$ & $10^2$ \\ \hline \hline
 %     SACLA repetition rate $[\mathrm{Hz}]$ & & 20 & 30 & 1.5 \\ 
 %     DAQ time [hour] & & 9& 34 & 3.8 \\ \hline
 %     number of pulses $[\mathrm{pulse}]$& &$6.5\times 10^{5}$  & $3.7\times 10^{6}$ & $5.7$ \\ \hline \hline
 %     horizontal beam width ($1\sigma$) $[\mathrm{\mu m}]$ & $\sigma_{\rm h}$  & 0.78 & 0.98& $0.8$ \\ \hline 
 %     vertical beam width ($1\sigma$) $[\mathrm{\mu m}]$ & $\sigma_{\rm v}$ & 207 & 134& $1.5$ \\ \hline
 %     detection efficiency for signals [\%] & $\epsilon$ & 12 & 13.2& $1.1$ \\ \hline
 %     method of upper-limit calculation [arb] & & 0.8 (conservative)& 1& $1.2$ \\ \hline \hline
 %     upper limit on cross section $[\mathrm{m^2}]$ & $\sigma_{\gamma\gamma\rightarrow\gamma\gamma}$ & $1.7 \times 10^{-24}$ & $1.9 \times 10^{-27}$ & $9\times 10^2$ \\ \hline 
 %   \end{tabular}
 %    \caption{A summary of enhancement factors compared to our previous experiment.}
 %   \label{tab:sum}
 % \end{center}
%\end{table}

\end{document}